\documentclass[10pt,journal,cspaper,compsoc]{IEEEtran}
\ifCLASSOPTIONcompsoc
   \usepackage[nocompress]{cite}
\else
\fi
\ifCLASSINFOpdf
\else
   \usepackage[dvips]{graphicx}
\fi
\usepackage[cmex10]{amsmath}
\usepackage{amssymb}
\usepackage{amsthm}

\newtheorem{thm}{Theorem}[section]     
\newtheorem{coro}{Corollary}[section]       
    
\newtheorem{lemm}{Lemma}[section]            
\newtheorem{defn}{Definition}[section]

\newtheorem{ex}{Example}[section]

\newtheorem{rem}{Remark}[section]

\newcommand{\mb}{\mathbf}
\newcommand{\mc}{\mathcal}
\newcommand{\he}{\mathcal H} % Entropia 
\newcommand{\info}{\mathcal I} %Informazione_Mutua
\DeclareMathOperator{\pr}{\textrm{P}} %definizione probabilità
\usepackage{tikz}
\usetikzlibrary{positioning,shapes,shadows,arrows,fadings}
\pgfdeclarelayer{nodelayer}
\pgfdeclarelayer{edgelayer}
\pgfsetlayers{nodelayer,edgelayer}
\usetikzlibrary{positioning,shapes,shadows,arrows,fadings}
\tikzstyle{user}=[circle,draw=gray!60,fill=gray!30,thick,scale=1.5, inner sep=0pt, minimum size=5mm]
\tikzstyle{userselected}=[circle,draw=red!130,fill=gray!30,thick,scale=1.5, inner sep=0pt, minimum size=5mm]
\tikzstyle{view}=[circle,draw=orange!100,fill=yellow!75,thick,scale=1.5, inner sep=0pt, minimum size=5mm]
\tikzstyle{view2}=[circle,draw=black!100,fill=red!50!,thick,scale=1.5, inner sep=0pt, minimum size=5mm]
\tikzstyle{arco}=[->,shorten >=1pt,>=stealth',semithick]

\begin{document}
\title{On the Equivalence of Two Security Notions\\
 for Hierarchical Key Assignment Schemes\\
  in the Unconditional Setting}

\author{\IEEEauthorblockN{Massimo~Cafaro,~\IEEEmembership{Senior~Member,~IEEE},
	Roberto~Civino,
        and~Barbara~Masucci}% <-this % stops an unwanted space
\IEEEcompsocitemizethanks{\IEEEcompsocthanksitem M. Cafaro is with the Department of Engineering for Innovation, University of Salento, Lecce 73100, Italy \protect\\     
E-mail: massimo.cafaro@unisalento.it
\IEEEcompsocthanksitem R. Civino is with the Department of Mathematics and Physics, University of Salento, Lecce 73100, Italy \protect\\    
E-mail:  roberto.civino@studenti.unisalento.it
\IEEEcompsocthanksitem B. Masucci is with the Department of Computer Science, University of Salerno, Fisciano 84084, Italy  \protect\\    
E-mail: masucci@dia.unisa.it}% <-this % stops a space
\thanks{}}


\markboth{Journal of \LaTeX\ Class Files,~Vol.~6, No.~1, January~2007}%
{Cafaro \MakeLowercase{\textit{et al.}}: On the Equivalence of Two Security Notions for Hierarchical Key Assignment Schemes in the Unconditional Setting}

%\IEEEpubid{\makebox[\columnwidth]{\hfill 0000--0000/00/\$00.00~\copyright~2007 IEEE}%

% PRIOR to the title within the \IEEEcompsoctitleabstractindextext IEEEtran
\IEEEcompsoctitleabstractindextext{%
\begin{abstract}
%\boldmath

The access control problem in a hierarchy can be solved by using a
\emph{hierarchical key assignment scheme}, where each class is
assigned an encryption key and some private information.
A formal security analysis for hierarchical key assignment  schemes has been traditionally 
considered in two different settings, i.e., the \emph{unconditionally secure}
and the \emph{computationally secure} setting, and with respect to two different notions:
security against key recovery  (KR-security) and security with respect to key indistinguishability (KI-security), with the latter notion being cryptographically stronger. 
Recently, Freire, Paterson and Poettering proposed \emph{strong key indistinguishability} (SKI-security) as a new security notion in the computationally secure setting, arguing that SKI-security is strictly stronger than KI-security in such a  setting.
In this paper we consider the unconditionally secure setting for hierarchical key assignment schemes.
In such a setting the security of the schemes is not based on 
specific unproven computational assumptions, i.e., it relies on the
theoretical impossibility of breaking them, despite the
computational power of an adversary  coalition.
We prove that, in this setting, SKI-security is not stronger than KI-security, i.e., the two notions are fully equivalent
from an information-theoretic point of view. 
\end{abstract}

% not want either math or citations to appear in the abstract.

% Note that keywords are not normally used for peer review papers.
}

\maketitle

% is not selected <OR> if conference mode is selected - because compsoc
% conference papers position the abstract like regular (non-compsoc)
\IEEEdisplaynotcompsoctitleabstractindextext

% \fi
\IEEEpeerreviewmaketitle

\section{Introduction}
% Computer Society journal papers do something a tad strange with the very
% first section heading (almost always called "Introduction"). They place it
% ABOVE the main text! IEEEtran.cls currently does not do this for you.
% However, You can achieve this effect by making LaTeX jump through some
% hoops via something like:
%
%\ifCLASSOPTIONcompsoc
%  \noindent\raisebox{2\baselineskip}[0pt][0pt]%
%  {\parbox{\columnwidth}{\section{Introduction}\label{sec:introduction}%
%  \global\everypar=\everypar}}%
%  \vspace{-1\baselineskip}\vspace{-\parskip}\par
%\else
%  \section{Introduction}\label{sec:introduction}\par
%\fi
%
% Admittedly, this is a hack and may well be fragile, but seems to do the
% trick for me. Note the need to keep any \label that may be used right
% after \section in the above as the hack puts \section within a raised box.


\IEEEPARstart{I}{n}  the \emph{access control} problem, we are concerned with the problem of ensuring that only authorized users of a computer system are entitled to access sensitive information, according to \emph{access control policies} that organize users in a hierarchy of disjoint classes,  called \emph{security classes}. The main reason for a hierarchy is the need to assign  different privileges and right sto users, according to their role, competencies and responsibilities. 

Hierarchies model and reflect many real world cases. A classical example is a hospital, in which doctors are allowed accessing data concerning their patients, but other people down in the hierarchy may access this information only partially (for instance, external researchers may be given access to anonymized data for clinical studies) or not at all (e.g., paramedics may be denied access).
Many different instances of the access control problem commonly arise in government agencies, in the military realm where classified documents can be accessed by selected users on the basis of their security clearance level, etc. 

In a hierarchical key assignment scheme, an encryption key and some private information are assigned  to each class in the hierarchy. The encryption key protects each class data by means of a  symmetric crypto-system, whereas  private information allows each class to compute the keys assigned to all of the classes lower down in the hierarchy. This assignment is carried out by a central authority, the so-called Trusted Authority (TA), which is active only in the initial distribution phase.

Hierarchical key assignment schemes were first proposed in the seminal paper of Akl and Taylor \cite{Akljournal}, and improved in order to achieve better performances or to allow hierarchies in which classes can be dynamically inserted or deleted \cite{Atallah}, \cite{harn}, \cite{hwang2}, \cite{wang}, \cite{lin2}, \cite{akl2}, \cite{sandhu}. 
Related work includes further research focusing on more general access control policies \cite{ipl}, \cite{lin}, \cite{ycn}. Even though a large number of schemes had been proposed, the lack of a formal security proof for many of them led to an initial analysis of the security properties of these schemes, which confirmed that, indeed, those schemes are insecure against collusive attacks in which a coalition of users tries attacking a class up in the hierarchy \cite{Chen}, \cite{shen}, \cite{Yeh}, \cite{ycn}. A useful taxonomy and evaluation of several schemes appeared in the literature can be found in \cite{crampton}.
A related branch of research, focusing on the design of hierarchical key assignment schemes with time-dependent keys,  called time-bound schemes, was started by Tzeng \cite{Tz}. However, Tzeng's scheme was shown later to be insecure against collusive attacks \cite{Yi}. Additional efforts include \cite{Chien}, \cite{huang}, \cite{Yeh}, but none of these schemes is secure against collusive attacks as well \cite{JoC}, \cite{DeFeMaTRep}, \cite{Tang}, \cite{Yi2}. 
New time-bound schemes, derived from the Akl-Taylor scheme,  were given by Wang and Laih \cite{WangLaih} and Tzeng \cite{Tzeng2} and further analyzed in \cite{mfcs09}.

The most used approach to hierarchical key assignment schemes is based on unproven specific computational assumptions.
In this computational setting, a hierarchical key assignment scheme is {\em provably-secure} under a complexity assumption if the existence of an adversary $A$ breaking the scheme is equivalent to the existence of an adversary $B$ breaking the computational assumption. The usual method of construction of
$B$ uses the adversary $A$ as a black-box. 
The need for formal security requirements in the computational setting was first addressed in \cite{Atallah} by Atallah et al., who  proposed two different security notions for hierarchical key assignment schemes: 
\emph{security against key recovery}  (KR-security) and \emph{security with respect to key indistinguishability} (KI-security).
KR-security corresponds to the requirement that an adversary is
not able to {\em compute} a key that it should not have access to. On the other hand, KI-security
formalizes the re\-qui\-rement that the adversary is not able {\em
to learn any information} about a key that it should not have access to, i.e., it is not able to distinguish it from a random
string having the same length.

Recently, Freire et al. \cite{Freire} proposed a new security notion for hierarchical key assignment schemes in the computationally secure setting.
Such a definition, called \emph{strong security with respect to key-indistinguishability} (SKI-security), formalizes the requirement
that the adversary is not able to learn any information  about a key
that it should not have access to, even if he has the additional capability of gaining access to encryption keys
associated to all of the classes above the target class in the hierarchy. Notice that these encryption keys might leak through usage and their compromise
could not directly lead to a compromise of the private information or the encryption key  of the target class.
Freire et al. \cite{Freire} argued that their new notion is strictly stronger than the existing KI-security notion, and provide SKI-secure constructions for
hierarchical key assignment schemes using pseudorandom functions and forward-secure pseudorandom generators. 
However, no formal security analysis was given  to prove that SKI-security is actually stronger than traditional KI-security.

In this paper we focus on an \emph{information-theoretic approach} which differs from the above computational approach since it does not depend on any computational assumption. Such an approach has already been considered in  \cite{ictcs}, \cite{dam} to analyze key assignment schemes for arbitrary access control policies, as well as in \cite{ipl13} to analyze time-bound hierarchical key assignment schemes.  
In such an \emph{unconditionally secure setting} the security of a hierarchical key assignment scheme  relies on the
theoretical impossibility of breaking it, despite the
computational power of an adversary  coalition.
In particular we prove that, in the unconditionally secure setting, SKI-security is not stronger than KI-security, i.e., the two definitions are, instead, fully equivalent from an information-theoretic point of view. 

The rest of this paper is organized as follows. 
In Section \ref{model} we first recall the formal definition of hierarchical key assignment schemes 
and formalize the notions of KI-security and SKI-security from an information-theoretic point of view. Then, we prove in Section \ref{equivalence} the equivalence, in the unconditionally secure setting, of the two security notions. Finally, we draw our conclusions  in Section \ref{conclusions}.

\section{Hierarchical Key Assignment Schemes}\label{model}
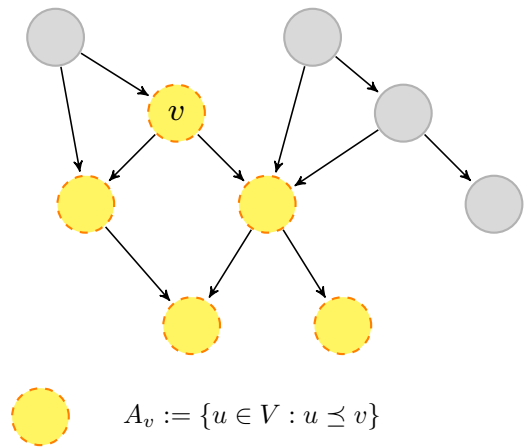
\begin{figure}[ht]
  \centering
  \begin{tikzpicture}[scale=0.8]
			\begin{pgfonlayer}{nodelayer}
				\node [style=view, dashed] (0) at (-2.25, 3) {\footnotesize $v$};
				\node [style=user] (1) at (1.5, 3) {};
				\node [style=view,dashed] (2) at (-0.75, 1.5) {};
				\node [style=view,dashed] (3) at (-3.75, 1.5) {};
				\node [style=user] (4) at (3, 1.5) {};
				\node [style=view,dashed] (5) at (0.5, -0.5) {};
				\node [style=view,dashed] (6) at (-2, -0.5) {};
				\node [style=user] (7) at (0, 4.25) {};
				\node [style=user] (8) at (-4.25, 4.25) {};
				\node [style=view, dashed] (9) at (-4.5, -2) {};
				\node [shape=rectangle] (10) at (-1, -2) {$A_v:=\{u \in V: u \preceq v\}$};
				\end{pgfonlayer}			
				\begin{pgfonlayer}{edgelayer}
				\draw [style=arco] (0) to (3);
				\draw [style=arco] (0) to (2);
				\draw [style=arco] (1) to (2);
				\draw [style=arco] (1) to (4);
				\draw [style=arco] (2) to (5);
				\draw [style=arco] (2) to (6);
				\draw [style=arco] (8) to (3);
				\draw [style=arco] (3) to (6);
				\draw [style=arco] (7) to (2);
				\draw [style=arco] (8) to (0);
				\draw [style=arco] (7) to (1);
		\end{pgfonlayer}
	\end{tikzpicture}  
	\caption{An example of the $A_v$ set.}
  \label{graph1}
\end{figure}

Consider a set of  users divided into a number of disjoint
classes, called {\em security classes}. A security class can
represent a person, a department or a user group in an
organization. A binary relation $\preceq$ that partially orders
the set of classes $V$ is defined in accordance with authority,
position or power of each class in $V$. The poset $(V,\preceq)$
is called a {\em partially ordered hierarchy}. For any two classes
$u$ and $v$, the notation $u\preceq v$ is used to indicate that
the users in $v$ can access $u$'s data.
We denote by $A_v$ the set ${\{u\in V: u\preceq v\}}$, for any $v\in V$, as shown in Figure \ref{graph1}.

The partially ordered hierarchy $(V,\preceq)$ can be
represented by a directed graph where each class
corresponds to a vertex in the graph and there is an edge from
class $v$ to class $u$ if and only if $u\preceq v$.
Further, this graph can be
simplified by eliminating all self-loops and edges which can be implied by
the property of the transitive closure. We denote by $\mc G=(V,E)$ the resulting
directed acyclic graph, which is called an  \emph{access graph}.

A {\em hierarchical key assignment scheme} for a partially ordered
hierarchy represented by an access graph $\mc G=(V,E)$  
is a method to assign a 
 private information $s_u$ and a key $k_u$ to each class $u\in V$. 
 The generation and distribution of the private
information and keys is carried out by a trusted third party, the
TA, which is connected to each class by means of a secure channel.
The encryption key $k_u$ can be used by users belonging to
class $u$ to protect their sensitive data by
means of a symmetric crypto-system, whereas, the private
information $s_v$ can be used by users belonging to
class $v$ to compute the key $k_u$ for any class $u\in A_v.$
%For each class $u\in V$,
%we denote by $S_{u}$ and $K_{u}$ the sets of all
%possible values that $s_{u}$ and $k_{u}$ can assume,
%respectively.

Following the line of \cite{ictcs,dam,ipl13}, we formally define hierarchical key assignment schemes by using the
entropy function (we refer the reader to the Appendix for some
properties of the entropy function and to \cite{Cover} for a
complete treatment of Information Theory), mainly because this
leads to a compact and simple description of the schemes and
because the entropy approach takes into account all of the probability
distributions on the keys assigned to the classes.
The same approach has been used in \cite{ictcs,dam,ipl13} to analyze different kinds of
key assignment schemes.

In the following, given a probability space $(\Omega, \mathcal F, \pr)$, we denote with a boldface
capital letter a random variable defined on $\Omega$ (e.g., $\mb X: \Omega \rightarrow \mathbb R$) and taking values on a set, denoted by the corresponding capital letter (e.g., $X\subseteq \mathbb R$). The values such a random variable can take are
denoted by the corresponding lower case letter (e.g., $x \in X$). Moreover, we write $p(x)$ for $p_{\mb X}(x):=\pr\{\mb X=x\}_{x \in X}$ to refer to the distribution of $\mb X$. Hereafter all of the random variables considered will be discrete. Given a random variable $\mb X$, we denote by $\he(\mb X)$ the Shannon entropy of $\mb X$.

Now we are ready to describe the {\em correctness} and {\em security} requirements that
a hierarchical key assignment scheme has to satifsy.

\medskip
\noindent{\bf Correctness Requirement:}
{\em  Each user can compute the key held by any class
lower down in the hierarchy}.\\
     Formally, for each class $v\in V$ and each class $u \in A_v$,  it holds that
$$\he(\mathbf{K}_u | \mathbf{S}_{v})=0.$$

\medskip
Notice that the correctness requirement is equivalent to saying
that the value of the private information $s_{v}$ held by
each user belonging to a class $v\in V$  corresponds to a unique value of the key
$k_{u}$, for each class $u\in A_v$.

\begin{defn}
\label{kas}
Let  $\mc G=(V,E)$ be an access graph. The set $\Sigma=\{(\mb K_u, \mb S_u)\}_{u \in V}$ is called a \emph{Hierarchical Key Assignment Scheme} for $\mc G$ if the random variables $\mb K_u$, $\mb S_u$, for each $u \in V$, satisfy the above correctness requirement.
\end{defn}

\begin{figure}[h]
  \centering
  \begin{tikzpicture}[scale=0.8]
			\begin{pgfonlayer}{nodelayer}
				\node [style=view2, loosely dashdotted] (0) at (-2.25, 3) {};
				\node [style=view2, loosely dashdotted] (1) at (1.5, 3) {};
				\node [style=userselected] (2) at (-0.75, 1.5) {\footnotesize $u$};
				\node [style=view, dashed] (3) at (-3.75, 1.5) {};
				\node [style=view,dashed] (4) at (3, 1.5) {};
				\node [style=view,dashed] (5) at (0.5, -0.5) {};
				\node [style=view,dashed] (6) at (-2, -0.5) {};
				\node [style=view2, loosely dashdotted] (7) at (0, 4.25) {};
				\node [style=view2, loosely dashdotted] (8) at (-4.25, 4.25) {};
				\node [style=view, dashed] (9) at (-4.5, -2) {};
				\node [shape=rectangle] (10) at (-0.75, -2) {$F_u:=\left\{v \in V: u \notin A_{v}\right\}$ };
				\node [style=view2, loosely dashdotted] (11) at (-4.5, -3.3) {};
				\node [shape=rectangle] (12) at (-0.11, -3.3) {${C_u:=\left\{v \in V: u \in A_v \right\}\setminus \{u\}}$ };
			\end{pgfonlayer}
			\begin{pgfonlayer}{edgelayer}
				\draw [style=arco] (0) to (3);
				\draw [style=arco] (0) to (2);
				\draw [style=arco] (1) to (2);
				\draw [style=arco] (1) to (4);
				\draw [style=arco] (2) to (5);
				\draw [style=arco] (2) to (6);
				\draw [style=arco] (8) to (3);
				\draw [style=arco] (3) to (6);
				\draw [style=arco] (7) to (2);
				\draw [style=arco] (8) to (0);
				\draw [style=arco] (7) to (1);
		\end{pgfonlayer}
	\end{tikzpicture}  
	\caption{An example of the $F_u$ and $C_u$ sets.}
  \label{graph2}
\end{figure}
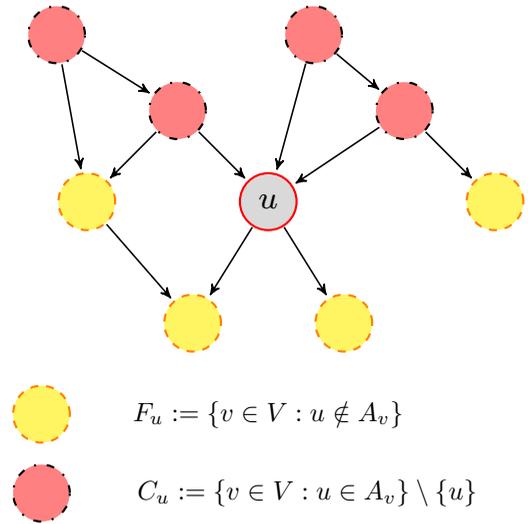

\medskip
In order to achieve security, given a class $u \in V$, the key $k_u$ should be protected against attacks mounted by a \emph{coalition} of users
belonging to each class $v$ such that $u\not\in A_v$. The amount of information available to the  coalition strictly depends on the kind of security one would like to attain: first, we  consider the case in which the coalition only owns the private information assigned to each class in the coalition; later, we will consider the case in which  the coalition also holds the encryption keys assigned to classes which can access class $u$.
Let ${F_{u}:=\{v \in V: u \not\in A_v\}}$ be the set of classes which are not allowed to access the sensitive data of users in class $u$.
Moreover,  let ${C_u:=\left\{v \in V: \quad u \in A_v \right\}\setminus \{u\}}$ be the set of classes, different from $u$, which are  
entitled to access the sensitive data of users in class $u$. Figure \ref{graph2} depicts the sets $F_{u}$ and $C_{u}$.
Regarding the relationship among the sets  $\{u\}$, $F_{u}$ and $C_{u\:}$, it can be easily verified that these sets are a \emph{partition} of 
the set of nodes $V$.

Given  $u \in V$ and $X \subseteq F_u$, where $X=\left\{v_1,v_2,\ldots,v_n\right\}$, let $\mb S_X$ be the random variable 
$\mb S_{v_1 : v_n} = (\mb S_{v_1}, \mb S_{v_2},\ldots, \mb S_{v_n})$  whose law is the joint law of the random variables $\mb S_{v_1}, \mb S_{v_2},\ldots, \mb S_{v_n}$. Similarly, given $Y \subseteq C_u$, 
where $Y=\left\{w_1,w_2,\ldots,w_m\right\}$, let $\mb K_Y$  be the random variable $\mb K_{w_1 : w_m} = (\mb K_{w_1}, \mb K_{w_2},\ldots, \mb K_{w_m})$
whose law is the joint law of the random variables $\mb K_{w_1}, \mb K_{w_2},\ldots, \mb K_{w_m}$.

We consider two different security notions: {\em security with respect to
key indistinguishability} (KI-security) and {\em strong security with respect to key
indistinguishability} (SKI-security). Security with respect to key indistinguishability
formalizes the re\-qui\-rement that the adversary coalition is not able {\em
to learn any information} about a key that it should not have
access to, i.e., it is not able to distinguish it from a random
string having the same length. 

\medskip
\noindent{\bf KI-Security Requirement:} {\em A coalition of users {\em have absolutely no information about} 
each key the coalition is not entitled to obtain.}\\
   Formally, for each class $u\in V$ and each coalition
$X\subseteq F_u$,
   it holds that  $$\he(\mb K_u | \mb S_X)=\he(\mb K_u).$$

   \begin{defn}
Let $\mc G=(V,E)$ be an access graph and let  $\Sigma=\{(\mb K_u, \mb S_u)\}_{u \in V}$ be
a hierarchical key assignment scheme for $\mc G$. The scheme $\Sigma$ \emph{provides key indistinguishability (or KI-security)} if the random variables $\mb K_u$, $\mb S_u$, for each $u \in V$, satisfy the above KI-security requirement.
\end{defn}

On the other hand, strong security with respect to key indistinguishability
formalizes the  re\-qui\-rement that the adversary coalition is not able 
to learn any information about a key that it should not have
access to, even though  the coalition has the additional capability of gaining access
to the encryption keys associated to other classes in the hierarchy.

\medskip
\noindent{\bf SKI-Security Requirement:} {\em A coalition of users {\em have absolutely no information about} 
each key the coalition is not entitled to obtain, even  though  the coalition has the additional capability of gaining access
to the encryption keys associated to other classes in the hierarchy.}\\
   Formally, for each class $u\in V$, each coalition
$X\subseteq F_u$, and each set  $Y\subseteq C_u$, it holds that
$$\he(\mb K_u | \mb S_X, \mb K_Y)=\he(\mb K_u).$$

   \begin{defn}
Let $\mc G=(V,E)$ be an access graph and let  $\Sigma=\{(\mb K_u, \mb S_u)\}_{u \in V}$ be
a hierarchical key assignment scheme for $\mc G$. The scheme $\Sigma$ \emph{provides strong key indistinguishability (or SKI-security)} if the random variables $\mb K_u$, $\mb S_u$, for each $u \in V$, satisfy the above SKI-security requirement.
\end{defn}

\begin{rem}
\label{cara}
It is evident that, given a scheme ${\Sigma=\{(\mb K_u, \mb S_u)\}_{u \in V}}$, in order to prove its KI-security it is enough proving that 
\begin{equation}
\forall u \in V,  \quad \he(\mb K_u | \mb S_{F_u})=\he(\mb K_u).
\label{caraKI}
\end{equation}

Eq. (\ref{caraKI}) and Lemma \ref{canthurtobs} grant that the requirement also holds for all subsets $X \subseteq F_u$. 
Analogously, in order to prove SKI-security  it is enough proving that  

\begin{equation}
\forall u \in V, \quad \he(\mb K_u | \mb S_{F_u}, \mb K_{C_u})=\he(\mb K_u).
\label{caraSKI}
\end{equation}
\end{rem}

\section{Equivalence of SKI-security and KI-security}
\label{equivalence}
In the following we study the relationship between the notions of KI-security and SKI-security for hierarchical key assignment schemes.
We first notice that SKI-security implies KI-security.

Indeed,  for each $u\in V$, each $X\subseteq F_u$, and each $Y\subseteq C_u$, it follows from Lemma \ref{canthurtobs} and \ref{canthurt} that
$$\he(\mb K_u | \mb S_X)\geq \he(\mb K_u | \mb S_X, \mb K_Y)=\he(\mb K_u).$$
Thus, if the SKI-security requirement holds, the KI-security requirement also holds. 

\medskip\noindent
We need the next definition.

\begin{defn}
Let $\mc G=(V,E)$ be an access graph and let $n \in \mathbb N$. The sequence $(u_1,u_2,\ldots,u_n) \in V^{n}$ is \emph{well ordered} if $n=1$ or $n>1$ and for each $j \in \{2,3,\ldots,n\}$ it holds $\left\{u_i\right\}_{i=1}^{j-1}\subseteq F_{u_j}$.
\end{defn}

We also recall the definition of {\em topological sorting} in a directed acyclic graph. 

\begin{defn}
Let $\mc G=(V,E)$ be a directed acyclic graph, let $n \in \mathbb N$, $V=\left\{u_i\right\}_{i=1}^{n}$ and let
$\sigma \in \mc S_n$ be a permutation. The sequence $X=(u_{\sigma(1)},u_{\sigma(2)},\ldots,u_{\sigma(n)})$ is a \emph{topological sort} of $\mc G$ if for each $u, v \in V$ such that $(u,v)\in E$, then $u$ appears before $v$ in $X$. 
\end{defn}

\begin{ex}
\label{ordtop}
Let $\mc G=(V,E)$ be an access graph, let $n \in \mathbb N$ and let $X=(u_n, u_{n-1}, \ldots, u_1)$ be a topological sort of $\mc G$.
It is easy to see that the sequence $(u_1, u_2, \ldots, u_n)$ is well ordered. Indeed, assume by contradiction that there  exist $i,j \in \{1,2,\ldots,n\}$, 
with $i \leq j$, such that the users belonging to  class $u_i$ \emph{are} authorized to access the information of class $u_j$.
Therefore we have that $(u_i, u_j)\in E$ and  it follows that $u_i$ appears before $u_j$ in $X$, so that $i>j$, which is a contradiction. 
\end{ex}

The following lemma is crucial to understand the relationship between KI-security and SKI-security.

\begin{lemm}
\label{lem}
Let $\mc G=(V,E)$ be an access graph, $\Sigma=\{(\mb K_u, \mb S_u)\}_{u \in V}$ be a KI-secure 
hierarchical key assignment scheme for $\mc G$ and let $n \in \mathbb N$.\linebreak  
If $(u_1,u_2,\ldots,u_n)\in V^{n}$ is a well ordered sequence, then 
$$\he(\mb K_{u_1:u_n}) = \sum_{j=1}^{n}{\he(\mb K_{u_j})}.$$
In particular, the random variables $\mb K_{u_1},\mb K_{u_2},\ldots,\mb K_{u_n}$ are independent. 
\end{lemm}

\begin{IEEEproof}
By Lemma \ref{indie} it is enough proving that
$$\he(\mb K_{u_1:u_n}) \geq \sum_{j=1}^{n}{\he(\mb K_{u_j})}.$$
Since $(u_1,u_2,\ldots,u_n)\in V^{n}$ is a well ordered sequence, for each $j \in \{2,3,\ldots,n\}$
it follows that $\left\{u_i\right\}_{i=1}^{j-1}\subseteq F_{u_j}$. 
Therefore, from the KI-security requirement we have that $$\he(\mb K_{u_j}|\mb S_{u_1:u_{j-1}})=\he(\mb K_{u_j}).$$ 
Moreover, from the correctness requirement, it follows that $$\he(\mb K_{u_j}|\mb S_{u_j})=0.$$
From Lemma \ref{refipl} it follows that

\begin{IEEEeqnarray*}{rCl}
\he(\mb K_{u_1 : u_{j-1}} | \mb S_{u_1 : u_{j-1}}) &\leq&\sum_{i=1}^{j-1}{\he(\mb K_{u_i} | \mb S_{u_1 : u_{j-1}})}\\
&\leq& \sum_{i=1}^{j-1}{\he(\mb K_{u_i} | \mb S_{u_i})}=0,
\end{IEEEeqnarray*}

\noindent
thus we have

\begin{equation}
\label{lem1}
\he(\mb K_{u_1:u_{j-1}} | \mb S_{u_1 : u_{j-1}})=0.
\end{equation}

\noindent
Moreover, by Lemma \ref{canthurtobs} it follows that
$$\he(\mb K_{u_1: u_{j-1}}| \mb S_{u_1 : u_{j-1}} ,\mb K_{u_j}) \leq \he(\mb K_{u_1 : u_{j-1}} | \mb S_{u_1 : u_{j-1}}),$$

\noindent
therefore,

\begin{equation}
\he(\mb K_{u_1 : u_{j-1}} | \mb S_{u_1 : u_{j-1}}, \mb K_{u_j})=0.
\label{lem2}
\end{equation}

\noindent
Consider the conditional mutual information \linebreak$\info\left(\mb K_{u_{j}};\mb K_{u_1 : u_{j-1}} | \mb S_{u_1 : u_{j-1}}\right)$. 
By using Lemma \ref{cmi} we deduce that

\begin{IEEEeqnarray*}{rCl}
&&\he(\mb K_{u_{j}} | \mb S_{u_1 : u_{j-1}})- \he(\mb K_{u_{j}} |\mb K_{u_1 : u_{j-1}}, \mb S_{u_1 : u_{j-1}})\\
&=& \he(\mb K_{u_1 : u_{j-1}} | \mb S_{u_1 : u_{j-1}})- \he(\mb K_{u_1 : u_{j-1}} | \mb K_{u_{j}} ,\mb S_{u_1 : u_{j-1}}),
\end{IEEEeqnarray*}

\noindent
and, by using Equations \eqref{lem1} e \eqref{lem2},  it follows that

\begin{equation*}
\he(\mb K_{u_{j}} | \mb S_{u_1 : u_{j-1}})= \he(\mb K_{u_{j}} |\mb K_{u_1 : u_{j-1}}, \mb S_{u_1 : u_{j-1}}).
\end{equation*}

\noindent
Therefore, from Lemma \ref{chainrule1} we have that

\begin{IEEEeqnarray*}{rCl}
\he(\mb K_{u_1:u_n}) &=& \he(\mb K_{u_1})+ \sum_{j=2}^{n}{\he(\mb K_{u_j} | \mb K_{u_1 : u_{j-1}})}\\
&\geq& \he(\mb K_{u_1})\\
&&+\sum_{j=2}^{n}{\he(\mb K_{u_j} | \mb K_{u_1 : u_{j-1}}, \mb S_{u_1 : u_{j-1}})}\\
&=& \he(\mb K_{u_1})+ \sum_{j=2}^{n}{\he(\mb K_{u_j} | \mb S_{u_1 : u_{j-1}})}\\
&=& \he(\mb K_{u_1})+ \sum_{j=2}^{n}{\he(\mb K_{u_j})}\\
&=& \sum_{j=1}^{n}{\he(\mb K_{u_j})}.
\end{IEEEeqnarray*}

\noindent
Moreover, from Lemma \ref{indie}, it holds that the random variables $\mb K_{u_1},\mb K_{u_2},\ldots,\mb K_{u_n}$ are independent. Thus, the lemma holds. 
\end{IEEEproof}

We are going to show that such a result allows proving the equivalence of the two security notions. Moreover, it is worth noting here that, reasoning as before, we can prove that KI-security is a sufficient condition for the independence of all of the keys in a hierarchical key assignment scheme. Indeed, the following theorem holds.

\begin{thm}
\label{kiindie}
Let $\mc G=(V,E)$ be an access graph and let $\Sigma=\{(\mb K_u, \mb S_u)\}_{u \in V}$ be a 
hierarchical key assignment scheme for $\mc G$. If $\Sigma$ is KI-secure for $\mc G$, then the random variables $\{\mb K_u\}_{u \in V}$ are independent.
\end{thm}

\begin{IEEEproof}
It suffices proceeding as in Lemma \ref{lem}, using the  well ordered sequence resulting by any arbitrary topological sort of $\mc G$ reverse ordered (see for instance Example \ref{ordtop}). 
\end{IEEEproof}

As said before, 
Lemma \ref{lem} will be used to prove the equivalence of KI-security and SKI-security. However, before proving the equivalence, we need some additional preliminary results.

\begin{lemm}
\label{lemmm}
Let $\mc G=(V,E)$ be an access graph, let $\Sigma=\{(\mb K_u, \mb S_u)\}_{u \in V}$ be a KI-secure 
hierarchical key assignment scheme for $\mc G$ and let $n,m \in \mathbb N$.\linebreak  If $(u_1,u_2,\ldots,u_{n+m})\in V^{n+m}$ is a well ordered sequence, then 
$$\he(\mb K_{u_{n+1} : u_{n+m}} | \mb K_{u_n}, \mb S_{u_1 : u_{n-1}}) = \sum_{j=1}^{m}{\he(\mb K_{u_{n+j}})}.$$
\end{lemm}

\begin{IEEEproof}
By Lemma \ref{canthurt} and \ref{indie}  it is enough showing that 
$$\he(\mb K_{u_{n+1} : u_{n+m}} |\mb K_{u_n}, \mb S_{u_1 : u_{n-1}}) \geq \sum_{j=1}^{m}{\he(\mb K_{u_{n+j}})}.$$
From the correctness requirement it follows that, for each $j \in \{2,3,\ldots, m+1\}$,

\begin{IEEEeqnarray*}{rCl}
\he(\mb K_{u_{n} : u_{n+j-1}} | \mb S_{u_1 : u_{n+j-1}}) &\leq& \sum_{i=0}^{j-1}{\he(\mb K_{u_{n+i}} | \mb S_{u_1 : u_{n+j-1}})}\\
&\leq& \sum_{i=0}^{j-1}{\he(\mb K_{u_{n+i}} | \mb S_{u_{n+i}})} =0,
\end{IEEEeqnarray*}

\noindent
thus
 
\begin{equation}
\he(\mb K_{u_{n} : u_{n+j-1}} | \mb S_{u_1 : u_{n+j-1}})=0,
\label{lemmm1}
\end{equation} 

\noindent
and consequently, 

\begin{equation}
\he(\mb K_{u_{n} : u_{n+j-1}} |\mb K_{u_{n+j}}, \mb S_{u_1 : u_{n+j-1}} )=0. 
\label{lemmm2}
\end{equation}

\noindent
Consider the conditional mutual information \linebreak$\info\left(\mb K_{u_{n+j}};\mb K_{u_{n} : u_{n+j-1}} | \mb S_{u_1 : u_{n+j-1}}\right)$. 
We have that

\begin{IEEEeqnarray*}{rCl}
&&\he(\mb K_{u_{n+j}} | \mb S_{u_1 : u_{n+j-1}})\\
&&- \he(\mb K_{u_{n+j}} |\mb K_{u_{n} : u_{n+j-1}}, \mb S_{u_1 : u_{n+j-1}})\\
&=& \he(\mb K_{u_{n} : u_{n+j-1}} | \mb S_{u_1 : u_{n+j-1}})\\
&&- \he(\mb K_{u_{n} : u_{n+j-1}} | \mb K_{u_{n+j}} ,\mb S_{u_1 : u_{n+j-1}}),
\end{IEEEeqnarray*}

\noindent
and, by using \eqref{lemmm1} and \eqref{lemmm2}, it follows that

\begin{equation*}
\he(\mb K_{u_{n+j}} | \mb S_{u_{1} : u_{n+j-1}})= \he(\mb K_{u_{n+j}} |\mb K_{u_{n} : u_{n+j-1}}, \mb S_{u_1 : u_{n+j-1}}).
\end{equation*}

\noindent
Now, letting $Q =  \he(\mb K_{u_{n+1} : u_{n+m}} |\mb K_{u_n}, \mb S_{u_1 : u_{n-1}})$, from Lemma \ref{chainrule2} we have that

\begin{IEEEeqnarray*}{rCl}
Q&=& \he(\mb K_{u_{n+1}} |\mb K_{u_n}, \mb S_{u_1 : u_{n-1}})\\
&&+ \sum_{j=2}^{m}{\he(\mb K_{u_{n+j}} |\mb K_{u_{n} : u_{n+j-1}}, \mb S_{u_1 : u_{n-1}})}\\
&\geq& \he(\mb K_{u_{n+1}} |\mb K_{u_n}, \mb S_{u_1 : u_{n}})\\&&+ \sum_{j=2}^{m}{\he(\mb K_{u_{n+j}} |\mb K_{u_{n} : u_{n+j-1}}, \mb S_{u_1 : u_{n+j-1}} )}\\
&=& \he(\mb K_{u_{n+1}} | \mb S_{u_1 : u_{n}})\\&&+ \sum_{j=2}^{m}{\he(\mb K_{u_{n+j}} | \mb S_{u_{1} : u_{n+j-1}})}\\
&=& \sum_{j=1}^{m}{\he(\mb K_{u_{n+j}})},
\end{IEEEeqnarray*}

\noindent
in which the last equality follows from the well ordering assumption and from the KI-security requirement.
Thus, the lemma holds.
\end{IEEEproof}

\begin{coro}
\label{lemm}
Let $\mc G=(V,E)$ be an access graph, let $\Sigma=\{(\mb K_u, \mb S_u)\}_{u \in V}$ be a KI-secure 
hierarchical key assignment scheme for $\mc G$ and let $n,m \in \mathbb N$.\linebreak  If  $(u_1,u_2,\ldots,u_{n+m})\in V^{n+m}$ is a well ordered sequence, then 
$$\he(\mb K_{u_{n+1} : u_{n+m}} | \mb S_{u_1 : u_{n-1}}) = \sum_{j=1}^{m}{\he(\mb K_{u_{n+j}})}.$$
\end{coro}

\begin{IEEEproof}
From Lemma \ref{canthurtobs} and \ref{lem} it follows that
 
\begin{IEEEeqnarray*}{rCl}
\he(\mb K_{u_{n+1} : u_{n+m}} | \mb K_{u_n}, \mb S_{u_1 : u_{n-1}}) &\leq& \he(\mb K_{u_{n+1} : u_{n+m}} | \mb S_{u_1 : u_{n-1}})\\
&\leq& \he(\mb K_{u_{n+1} : u_{n+m}})\\
&=& \sum_{j=1}^{m}{\he(\mb K_{u_{n+j}})},
\end{IEEEeqnarray*}

\noindent
consequently, the thesis follows from Lemma \ref{lemmm}.
\end{IEEEproof}

\begin{thm}
\label{kiski}
Let $\mc G=(V,E)$ be an access graph, ${\Sigma=\{(\mb K_u, \mb S_u)\}_{u \in V}}$ a KI-secure hierarchical key assignment scheme for $\mc G$, $n,m \in \mathbb N$.\linebreak  If $(u_1,u_2,\ldots,u_{n+m})\in V^{n+m}$ s a well ordered sequence, then  
$$\he(\mb K_{u_n} | \mb K_{u_{n+1} : u_{n+m}}, \mb S_{u_1 : u_{n-1}})=\he(\mb K_{u_n}).$$
\end{thm}

\begin{IEEEproof}
Consider the conditional mutual information $\info(\mb K_{u_n};\mb K_{u_{n+1} : u_{n+m}} | \mb S_{u_1 : u_{n-1}})$. It holds that

\begin{IEEEeqnarray}{rCl}
&&\he(\mb K_{u_n} | \mb S_{u_1 : u_{n-1}})- \he(\mb K_{u_n} | \mb K_{u_{n+1} : u_{n+m}}, \mb S_{u_1 : u_{n-1}})\nonumber \\
&=&\he(\mb K_{u_{n+1} : u_{n+m}} | \mb S_{u_1 : u_{n-1}})\nonumber \\
&&- \he( \mb K_{u_{n+1} : u_{n+m}} | \mb K_{u_n}, \mb S_{u_1 : u_{n-1}}) \nonumber \\
&=&\sum_{j=1}^{m}{\he(\mb K_{u_{n+j}})}-\sum_{j=1}^{m}{\he(\mb K_{u_{n+j}})}\label{teo1}\\
&=& 0 \nonumber
\end{IEEEeqnarray}

\noindent
where in Eq. (\ref{teo1}) we used Corollary \ref{lemm} and Lemma \ref{lemmm}. 
It holds that 
\begin{IEEEeqnarray*}{rCl}
\he(\mb K_{u_n} | \mb K_{u_{n+1} : u_{n+m}}, \mb S_{u_1 : u_{n-1}}) &=& \he(\mb K_{u_n} | \mb S_{u_1 : u_{n-1}}) \\
&=&\he(\mb K_ {u_n}),
\end{IEEEeqnarray*}

\noindent
where in the last equality we used the KI-security requirement. Thus, the theorem holds.
\end{IEEEproof}

We are now in the position to state the following theorem, which is the main result and contribution of this paper.

\begin{thm}
Let $\mc G=(V,E)$ be an access graph and let $\Sigma=\{(\mb K_u, \mb S_u)\}_{u \in V}$ be a 
hierarchical key assignment scheme for $\mc G$. If $\Sigma$ is KI-secure for $\mc G$, then $\Sigma$ is also SKI-secure for $\mc G$. 
\end{thm}

\begin{IEEEproof}
Let $u \in V$. By Remark \ref{cara}, the SKI-security of $\Sigma$ is proved if we can show that Eq. (\ref{caraSKI}) holds, i.e., $$\he(\mb K_u | \mb K_{C_u}, \mb S_{F_u}) = \he(\mb K_u).$$

Let ${n, m \in \mathbb N}$ and ${u_1, u_2,\ldots, u_{n+m}}$ be an 
enumeration of $V$ such that ${\{u_i\}_{i=1}^{n-1}=F_u}$,  ${u_n=u}$,  and ${\{u_i\}_{i=n+1}^{n+m}=C_u}$. Moreover, let ${\sigma \in \mc S_{n-1}}$ and ${\tau \in \mc S_{m}}$ be permutations such that ${(u_{\sigma(1)}, u_{\sigma(2)}, \ldots, u_{\sigma(n-1)})}$ is a topological sort of $F_u$ and ${(u_{\tau(n+1)}, u_{\tau(n+2)}, \ldots, u_{\tau(n+m)})}$ a topological sort of $C_u$. Then, the sequence
\small
$$(u_{\sigma(n-1)}, \ldots, u_{\sigma(2)}, u_{\sigma(1)}, u_n, u_{\tau(n+m)}, \ldots, u_{\tau(n+2)}, u_{\tau(n+1)})$$
\normalsize

\noindent
is well ordered, thus, by Theorem \ref{kiski} it follows that
$$\he(\mb K_{u_n} | \mb K_{u_{\tau(n+m)} : u_{\tau(n+1)}}, \mb S_{u_{\sigma(n-1)} : u_{\sigma(1)}}) = \he(\mb K_{u_n}),$$ i.e.,  $$\he(\mb K_u | \mb K_{C_u}, \mb S_{F_u}) = \he(\mb K_u).$$  
Thus, the theorem holds.
\end{IEEEproof}

\section{Conclusions}
\label{conclusions}
Freire, Paterson and Poettering \cite{Freire} recently proposed a new security notion for hierarchical key assignment schemes in the computationally secure setting, called \emph{strong security with respect to key indistinguishability} (SKI-security). They argued that their new notion is stronger than the traditional KI-security one, but did not prove formally that indeed this is the case. In this paper we showed that, in the unconditionally secure setting, SKI-security is not stronger than KI-security, i.e., the two definitions are fully equivalent from an information-theoretic point of view. 

\appendices

\section{}
We recall here a few basic results of Information Theory used in our definitions and proofs. For a
complete treatment of the subject the reader is advised to consult \cite{Cover}.

\begin{defn}
Let $\mb X$ be a random variable with distribution $p(x)$. The \emph{entropy} of $\mb X$ is the operator 
$$\he(\mb X):=-\sum_{x\in X}{p(x)\log(p(x))},$$
where $\log$ denotes base 2 logarithm.
\end{defn}

\noindent
The entropy satisfies the following property

$$\he(\mb X) \geq 0$$ with equality if and only if there exists $x \in X$ such that $p(x) = 1$.

\begin{defn}
Let $\mb X, \mb Y$ be random variables with joint distribution $p(x,y)$. The \emph{joint entropy} of $\mb X$ and $\mb Y$ is the operator $$\he(\mb X, \mb Y):=-\sum_{x\in X}\sum_{y \in Y}{p(x,y)\log\left(p(x,y)\right)}.$$
\end{defn}

\begin{defn}
Let $\mb X, \mb Y$ be random variables with joint distribution $p(x,y)$ and conditional distribution $p(x | y)$. The \emph{conditional entropy} of $\mb X$ given $\mb Y$ is the operator 
$$\he(\mb X | \mb Y):=-\sum_{x \in X}\sum_{y \in Y}{p(x,y)\log(p(x | y))}.$$
\end{defn}

\begin{lemm}
\label{canthurt}
Let $\mb X, \mb Y$ be random variables. Then, $$\he(\mb X | \mb Y) \leq \he(\mb X).$$
Moreover, the equality holds if and only if $\mb X$ e $\mb Y$ are independent. 
\end{lemm}

\begin{lemm} 
\label{indie}
Let $n \in \mathbb N$ and let $\mb X_1, \mb X_2,\ldots, \mb X_n$ be random variables. It holds that
$$\he(\mb X_{1:n})\leq \sum_{i=1}^{n}{\he(\mb X_i)}.$$
Moreover, the equality holds if and only if $\mb X_1, \mb X_2,\ldots, \mb X_n$ are independent. 
\end{lemm} 

\begin{lemm} 
\label{refipl}
Let $n \in \mathbb N$ and let $\mb X_1, \mb X_2,\ldots, \mb X_n, \mb Y$ be random variables. It holds that
$$\he(\mb X_{1:n}| \mb Y)\leq \sum_{i=1}^{n}{\he(\mb X_i | \mb Y)}.$$ 
\end{lemm} 

\begin{lemm}
\label{chainrule1}
Let $n \in \mathbb N\setminus \{1\}$ and let $\mb X_1, \mb X_2,\ldots, \mb X_n$ be random variables. It holds that
$$\he(\mb X_{1,n})=\he(\mb X_1)+\sum_{i=2}^{n}{\he(\mb X_i | \mb X_{1:{(i-1)}})}.$$
\end{lemm}

\begin{lemm}
\label{chainrule2}
Let $n \in \mathbb N\setminus \{1\}$ and let $\mb X_1, \mb X_2,\ldots, \mb X_n, \mb Y$ be random variables. It holds that
$$\he(\mb X_{1:n} | \mb Y)=\he(\mb X_1 | \mb Y)+\sum_{i=2}^{n}{\he(\mb X_i | \mb X_{1:(i-1)}, \mb Y)}.$$
\end{lemm}

\begin{defn}
Let $\mb X, \mb Y$ be random variables. The \emph{mutual information} between $\mb X$ and $\mb Y$ is the operator
$$\info(\mb X; \mb Y) := \he(\mb X) - \he(\mb X | \mb Y).$$
\end{defn}

From Lemma \ref{canthurt} follows that $\info(\mb X; \mb Y) \geq 0$ and it is easy to verify that $\info(\mb X; \mb Y) = \info(\mb Y; \mb X)$.

\begin{defn}
Let $\mb X, \mb Y, \mb Z$ be random variables. The \emph{conditional mutual information} of $\mb X$ and $\mb Y$ given $\mb Z$ is the operator $$\info(\mb X; \mb Y | \mb Z):= \he(\mb X | \mb Z)-\he(\mb X | \mb Y, \mb Z).$$
\end{defn}

\begin{lemm} 
\label{canthurtobs}
Let $\mb X, \mb Y, \mb Z$ be random variables. It holds that  $\info(\mb X; \mb Y |\mb Z)\geq 0.$ Moreover, $$\he(\mb X | \mb Y, \mb Z) \leq \he(\mb X | \mb Z).$$
\end{lemm} 

\begin{lemm} 
\label{cmi}
Let $\mb X, \mb Y, \mb Z$ be random variables. It holds that
$$\info(\mb X; \mb Y | \mb Z)=\info(\mb Y; \mb X | \mb Z).$$  In particular,
$$\he(\mb X | \mb Z)-\he(\mb X | \mb Y, \mb Z)= \he(\mb Y | \mb Z)-\he(\mb Y | \mb X, \mb Z).$$
\end{lemm}

\bibliographystyle{abbrv}
\bibliography{bibliography}


\begin{IEEEbiography}[{\includegraphics[width=1in,height=1.25in,clip,keepaspectratio]{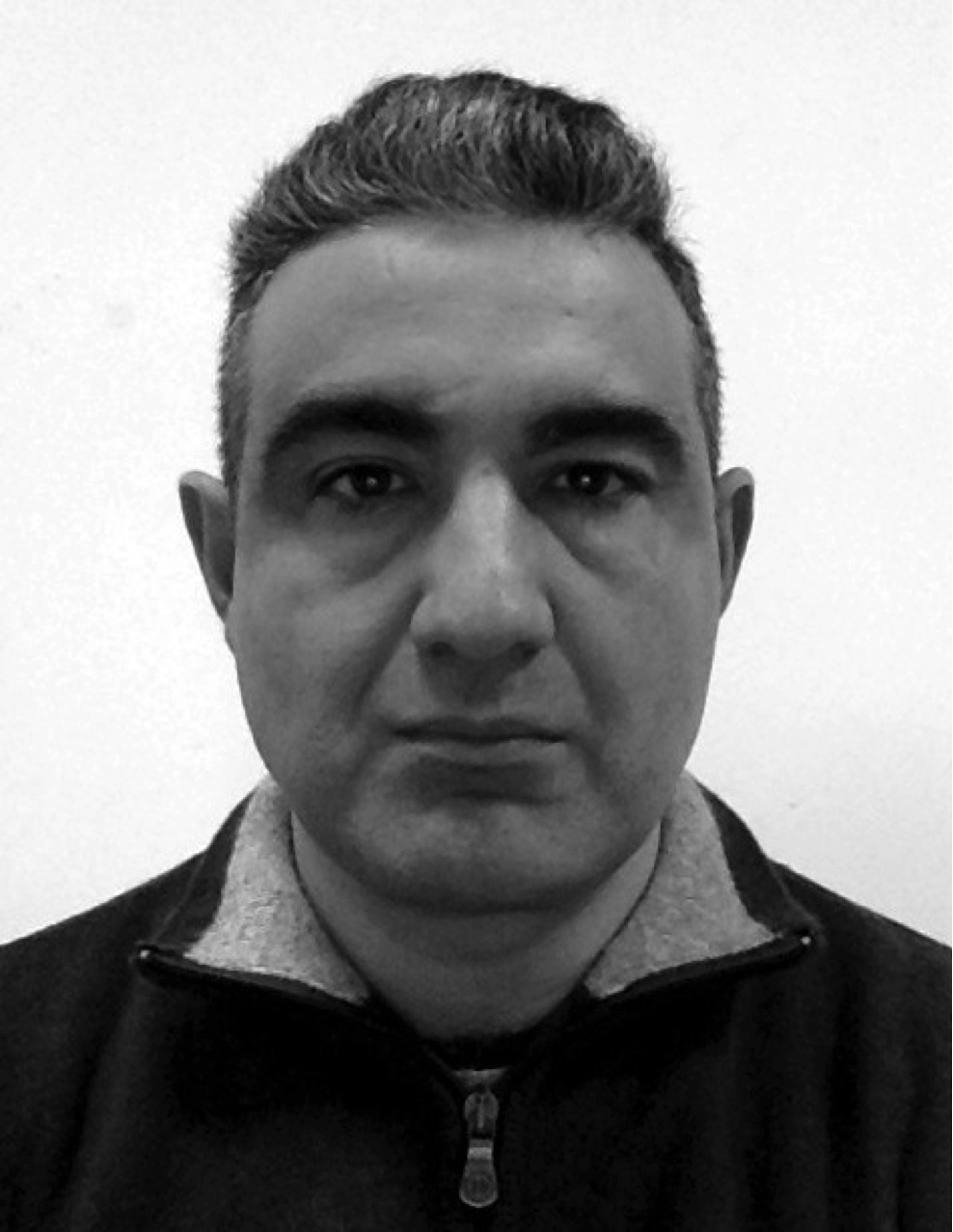}}]{Massimo Cafaro}
is an Assistant Professor at the Department of Innovation Engineering of the University of Salento. His research covers High Performance, Distributed and Cloud/Grid Computing, security and cryptography. He received a degree in Computer Science from the University of Salerno and a Ph.D. in Computer Science from the University of Bari. He is a Senior Member of IEEE and of IEEE Computer Society, and Senior Member ACM. He authored or co-authored more than 90 refereed papers on parallel, distributed and grid/cloud computing. He co-authored and holds a patent on distributed database technologies. His research interests are focused on both theoretical and practical aspects of parallel and distributed computing, security and cryptography, with particular attention to the design and analysis of algorithms.
\end{IEEEbiography}

\begin{IEEEbiography}[{\includegraphics[width=1in,height=1.25in,clip,keepaspectratio]{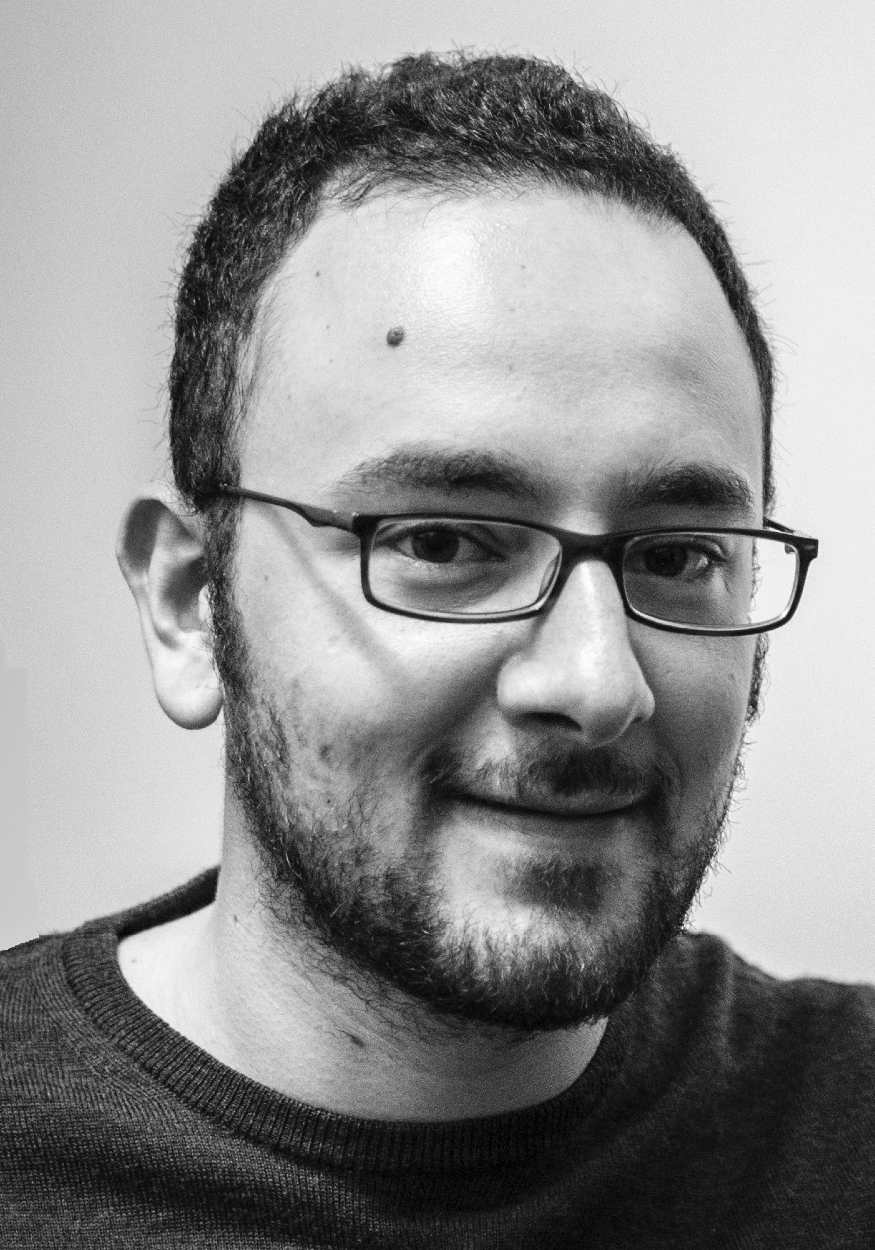}}]{Roberto Civino}
received the M.Sc. degree in Mathematics (cum laude) from the University of Salento in 2014. His interests are in the field of security and cryptography.
\end{IEEEbiography}

\begin{biography}
[{\includegraphics[width=1in,height=1.25in,clip,keepaspectratio]
{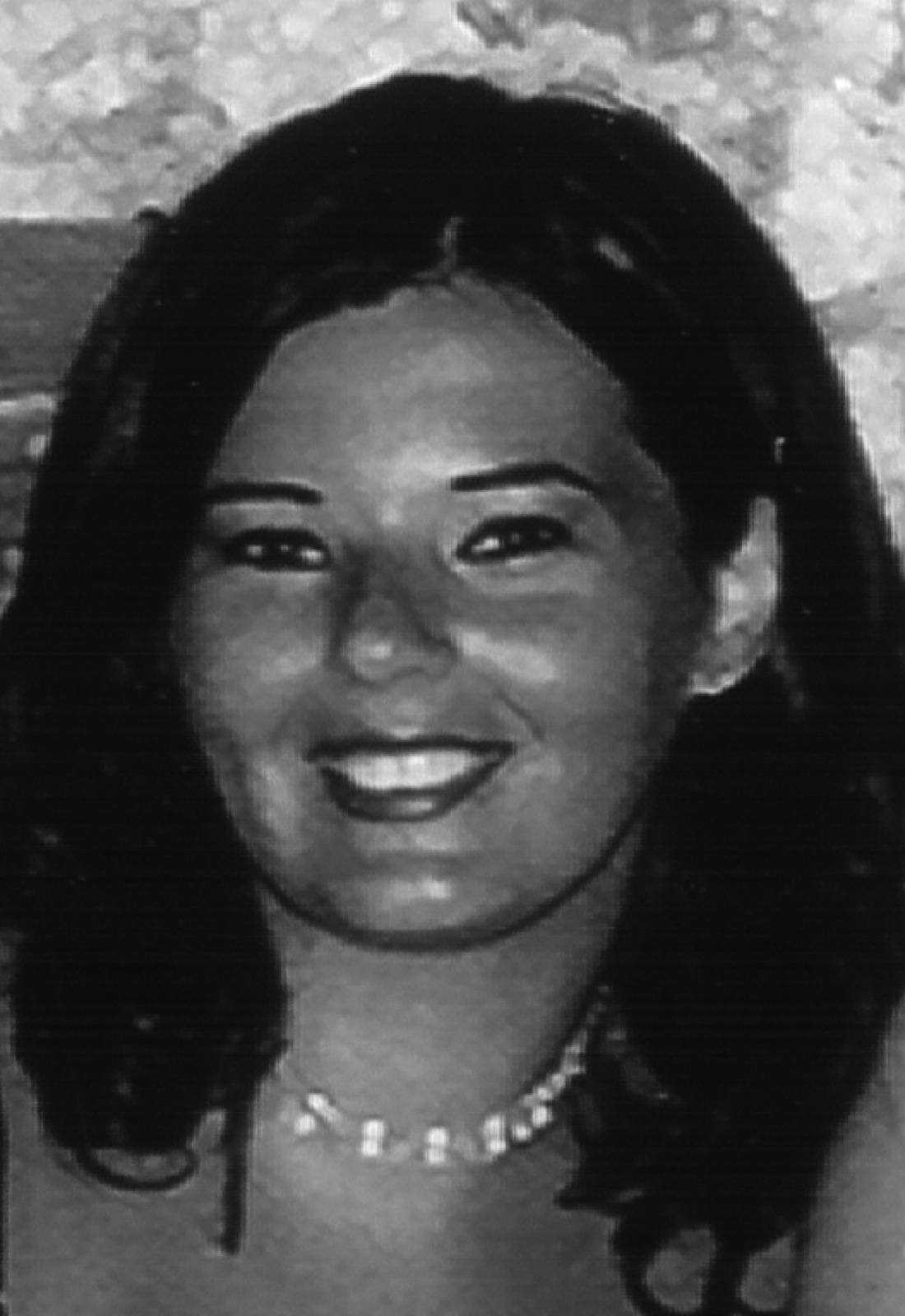}}]{Barbara Masucci}
 received the Laurea degree in Computer Science (cum laude) 
from the University of Salerno in 1996.
From September 1999 to April 2000 she was a Visiting 
Researcher at the Centre for Applied Cryptographic Research, 
in the Department of Combinatorics and Optimization 
of the University of Waterloo, Ontario, Canada.
In 2001 she received a Ph.D.
in Computer Science from the University of Salerno. 
From March 2001 to February 2002 she was a Post-Doctoral Fellow
at the Dipartimento di Informatica ed Applicazioni,
University of Salerno.
In July 2001 she visited the Departament de 
Matem\'atica Aplicada IV at the  Universitat Polit\'ecnica de Catalunya, 
Barcelona, Spain. 
Since March 2002 she has been an Assistant Professor at 
the Dipartimento di Informatica,
University of Salerno.
Her research interests include Algorithms, Data Security, and Cryptography. 
In particular, she has been working on 
the design and analysis of secure and efficient
cryptographic protocols. 

\end{biography}
\end{document}